\documentstyle[12pt,epsfig]{article}  

\newcounter{multieqs}

\newcommand{\bq}{\begin{equation}}
\newcommand{\fq}{\end{equation}}
\newcommand{\bqr}{\begin{eqnarray}}
\newcommand{\fqr}{\end{eqnarray}}

\newcommand{\rf}[1]{(\ref{#1})}

 \def\cN{{\cal N}}

\def\pr{^{\prime}}

\begin{document}

\thispagestyle{empty}

\marginparwidth = .5in

\marginparsep = 1.2in

\begin{flushright}
\begin{tabular}{l}
ANL-HEP-PR-00-83 \\

hep-th/0008162
\end{tabular}
\end{flushright}

\vspace{40mm}
\begin{center}

{\Large\bf On the Finiteness of $\cN=8$ Quantum Supergravity}\\  

\vspace{14mm}

{\bf Gordon Chalmers}
\\[5mm]
{\em Argonne National Laboratory \\
High Energy Physics Division \\
9700 South Cass Avenue \\
Argonne, IL  60439-4815 } \\[5mm]  

\vspace{12mm}

{\bf Abstract}   

\end{center}

We describe the constraints imposed on quantum maximal supergravity theories in 
perturbation theory by the two duality frameworks: S-duality in superstring 
theory and the AdS/CFT holographic correspondence between IIB superstring theory 
and $\cN=4$ super Yang-Mills theory.  

\vfill 
\line(6,0){220} \vskip .04in
E-mail address: chalmers@pcl9.hep.anl.gov \hfill

\setcounter{page}{0}
\newpage
\setcounter{footnote}{0}

\baselineskip=16pt

\section{Introduction}  

Constructing a theory of quantum gravity has motivated the formulation of both 
supergravity theories \cite{Freedman:1976py,Freedman:1976xh} and superstring 
theories.  Developments in the non-perturbative duality structures in 
superstring theories, in the art of calculating supergravity amplitudes, and 
in the formulation of a string theory description of $\cN=4$ super Yang-Mills 
theory have allowed a fresh look at supergravity theories in different dimensions.  
In particular, maximal supergravities (preserving $\cN=32$ supersymmetries) constructed in
\cite{deWit:1977fk,Cremmer:1978km,Cremmer:1978ds,deWit:1982ig} are re-examined in 
this context.  In the past, and in the absence of any calculation ever producing a 
divergence in the maximally supersymmetric gravitational theory, $\cN=8$ supergravity 
was superseded by the emergence of superstring theory, the non-perturbative form 
of which is presumably better defined than the former theory due to singularities. 
A counterterm compatible with the full $\cN=8$ supersymmetry was initially constructed 
in \cite{Howe:1981th,Kallosh:1981fi,Howe:1981xy}.  

Gravity theories in general dimensions have a dimensionful coupling constant 
that precludes the notion of perturbative renormalizability in the sense of 
Wilson.  Duality in field and string theory predicts structures, however, in 
the coupling constants beyond what natural renormalization criteria suggests.  
Furthermore, duality structures are potentially leading to exact solutions of 
higher-dimensional theories, notably the large $N_c$ limit of $\cN=4$ super 
Yang-Mills gauge theory via tree-level scattering in IIB superstring theory.  
Finite coupling structure, or the intermediate coupling regime, is a corner of 
coupling space that is not related to the free-field limit by an inverse 
transformation in an infinitely large coupling constant in general, and these 
conclusions are based on investigations of this region in the parameter space.  

In this letter we summarize the developments and predictions for the finiteness 
properties of maximally extended supergravity theories that have arisen through 
S-duality compatible graviton scattering\footnote{The quantum IIB superstring theory 
is conjectured to be self-dual under fractional linear transformations of the coupling 
constant $\tau= \chi+ie^{-\phi}$.  Although the truncation to the massless modes clearly 
does not remain S-duality invariant at the quantum level as can be found by explicit 
calculations, in contrast to the field equations of the supergravity, there is a 
remnant of the superstring structure that persists in the perturbative sector.} 
and in the holographic mapping between IIB superstring theory on anti-de Sitter 
space-times and boundary conformal field theories.  

\section{S- and U-duality in perturbation theory} 
\setcounter{equation}{0}

The scattering of gravitons in Einstein frame, in accord with U-duality of IIB 
superstring (and M-) theory and to infinite genus order, must be 
invariant under the non-perturbative  transformation of the string coupling constant.  
The form and the implications for $\cN=8$ supergravity have been explored in
\cite{Chalmers:2000zg}.   The ten-dimensional form of the graviton scattering in 
Einstein frame has the form within the derivative expansion, 
\bqr  
S_{4-pt}^{\rm IIB} = \frac{1}{\alpha'}
\sum_{k=0}^\infty \int d^{10}x \sqrt{g} ~ 
\alpha'{}^k \,  f_k(\tau,\bar\tau) \Box^k R^4 \ \, , 
\label{sform1}
\fqr 
together with a series of non-analytic terms which may be reconstructed in 
a manifestly S-duality invariant fashion through unitarity relations leading to 
unitary scattering; the tensor structure associated with the placement of the 
derivatives $\Box^k$ in \rf{sform1} is implied and not relevant for the analysis 
in this work although certainly so for a determination of the complete four-point 
scattering.  The functions $f_k(\tau,\bar\tau)$ generically have an expansion in 
accord with the dilaton in perturbative string theory (and quantum supergravity) of 
\bqr 
f_k(\tau,\bar\tau) = a_0^{(k)} \tau_2^{{3\over 2}+{k\over 2}} 
+ a_1^{(k)} \tau_2^{-{1\over 2}+{k\over 2}} + a_2^{(k)} 
 \tau_2^{-{5\over 2}+{k\over 2}} + \ldots \ . 
\label{form}
\fqr 
The tensor $R^4$ is eight-derivatives and is an integral over all of on-shell 
constrained IIB superspace; its appearance in IIB string theory is well-known 
\cite{Green:1982sw} and its tensor structure is a consequence of maximal supersymmetry.  
On the fundamental domain of $U(1)\backslash SL(2,R)/SL(2,Z)$, which labels the 
vacua of uncompactified IIB superstring theory and is parameterized by the space 
of coupling constants,   
\bqr  
{\cal F}_1 = \Bigl\{ \tau=\tau_1+ i\tau_2 
  : \tau_1^2+\tau_2^2 \geq 1 ~,~ \vert \tau_1\vert \leq {1\over 2} 
 \Bigr\}  \ ,  
\label{modregion} 
\fqr 
any function invariant under $SL(2,Z)$ may be decomposed on 
the product of the set of functions, 
\bqr  
E_s^{(q,-q)} (\tau,\bar\tau) = \sum_{(m,n)\neq (0,0)} {\tau_2^s\over 
 (m+n\tau)^{s-q} (m+n\bar\tau)^{s+q}}  \ , 
\label{eisenstein}
\fqr 
where $\sum_i q_i=0$ in the factors $E^{(q_i,-q_i)}_{s_i}$ (as in \cite{Chalmers:2000zg}), 
with $s=\sum_j s_j$, and with $s\geq 3/2$ or $s={1\over 2}+it$ together with cusp forms 
possessing the expansion on the fundamental domain, 
\bqr  
f_{\rm cusp}(\tau,\bar\tau) = \sum_{n\neq 0} a_n \tau_2^{1\over 2} K_{n-1/2} 
(2\pi \vert n\vert \tau_2) e^{2\pi i n\tau_1} \ ,
\label{cusp}
\fqr 
with undetermined (but bounded) coefficients $a_n$ \cite{Terras} (no explicit 
examples of cusp forms are known and string theory may produce examples in the 
scattering).  Eisenstein functions have appeared originally in the modular 
construction of the $R^4$ term involving D-instanton contributions \cite{Green:1997tv}
and in the context involving $(p,q)$ string-instantons in \cite{Kiritsis:1997em} (see also 
\cite{Obers:2000um} in more general cases).  For large values of $\tau_2$ the 
modular functions $E_s(\tau,{\bar\tau})$ functions have the expansion, following 
a Poisson resummation, 
\bqr 
& &  
E_s(\tau,\bar\tau) = 2\zeta(2s) \tau_2^s + 2\sqrt{\pi} \zeta(2s-1) 
{\Gamma(s-{1\over 2})\over\Gamma(s)} \tau_2^{1-s}  
\cr && \hskip .4in 
+ {2\sqrt{\tau_2} \pi^s\over \Gamma(s)} \sum_{n\geq 1,m\neq 0} 
\vert{m\over n}\vert^{s-{1\over 2}} 
K_{s-{1\over 2}}(2\pi \vert m n\vert \tau_2) e^{2\pi i m n\tau_1} \ ,  
\fqr 
with $K_a$ the standard modified Bessel function.  
According to the Roelcke-Selberg decomposition \cite{Terras} on the fundamental 
domain \rf{modregion} any normalizable $SL(2,Z)$ function (with integration 
measure $d^2\tau/\tau_2^2$) can be written as 
\bqr  
f(\tau,\bar\tau) = \sum_{n\geq 0} f_{cusp}^{(n)}(\tau,\bar\tau) \langle 
f_{\rm cusp}^{(n)}, f\rangle + {1\over 4\pi i} \int_{t={1\over 2}} dt~ 
E_{{1\over 2}+it} \langle E_{{1\over 2}+it}, f\rangle \ ,  
\label{decomposition}
\fqr 
with the integration along the imaginary axis and $\langle ...\rangle$ denoting an 
integration with measure $d^2\tau/\tau_2^2$.  The difference between a normalizable 
and non-normalizable function is contained in the covariantization of the non-normalizable 
terms proportional to $\tau_2^s$ with $s\geq 3/2$; these covariantizations are listed 
in \rf{eisenstein}.\footnote{Further modular invariant functions may be deduced by enforcing 
the values of functions to be equal on the images of the fundamental domain under the 
fractional linear transformations.  These examples, however, are singular and are 
excluded under the demand that the dilaton dependence of the string scattering is smooth 
on the fundamental domain.  I thank E. Martinec for a discussion on this point.} 

The functions in \rf{cusp}\footnote{An explicit example of a cusp form has not  
been constructed to date, and string scattering may lead to examples in the non-perturbative 
regime.} have an expansion in powers of exponentials of the coupling constant 
and model D-instanton corrections to the amplitudes in the ten-dimensional case; 
these functions in \rf{cusp} do not contribute to the perturbative expansion of 
either the superstring or in the supergravity limit.  The set of functions in 
\rf{eisenstein} and \rf{cusp} span a basis on which all smooth $SL(2,Z)$ invariant 
functions may be expanded upon (for normalizable functions this is examined in 
\cite{Terras}).  The latter two sets of functions are $L^2$ normalizable on the 
fundamental domain, and the remaining (products of functions in \rf{eisenstein} 
with $\sum_i q_i=0$) parameterize the $SL(2,Z)$ completion of the 
divergent pieces of a general $SL(2,Z)$ invariant function.  This set of functions 
spans the space of the coefficents of the derivative expansion of the $SL(2,Z)$ 
invariant graviton scattering amplitude.  This completes our summary of the general
form of S-duality compliant graviton scattering previous to including further 
constraints; the functions $E_{1/2+it}$ individually have the large $\tau_2$ expansion 
of $\tau_2^{1/2}\cos(\tau_2)$; potentially their contribution in \rf{decomposition} is 
compatible with the perturbative structure of string scattering through the transform 
in \rf{decomposition} but the non-perturbative instantonic corrections are not generally.  

Furthermore, the scattering has been computed from the string up to genus two in the 
string-inspired supergravity approximation and twelve derivatives.  We now turn to 
implementing the modular construction to $\cN=4$ super Yang-Mills theory via the AdS/CFT 
correspondence at all values of the coupling constant.  

\section{Graviton scattering and AdS/CFT} 
\setcounter{equation}{0}

The holographic correspondence between IIB superstring theory and $\cN=4$ 
super Yang-Mills theory \cite{Maldacena:1998re,Gubser:1998bc,Witten:1998qj} is 
specified in part by the couplings of the $SU(N_c)$ gauge theory $(\tau_{\rm YM} 
= {\theta_{YM} \over 2\pi} + i{4\pi\over g_{YM}^2}, N_c)$ and string theory 
$(\tau={\theta \over 2\pi}+{i\over g_s},\alpha'$) identifications 
\cite{Maldacena:1998re}   
\bqr  
{R_{AdS}^2\over \alpha\pr} = \lambda \ , \qquad 4\pi g_s = g^2_{\rm YM} \ , 
\qquad \theta_s=\theta_{YM}
\label{couplings}
\fqr 
together with propagating string theory \cite{Witten:1998qj} on the $AdS_5\times S^5$
background with radius $R_{AdS}$ and a non-vanishing five-form field flux in 
the $AdS_5$ directions,  
\bqr  
F_{\mu_1\ldots\mu_5}= {1\over R_{AdS}} \epsilon_{\mu_1\ldots\mu_5} \qquad 
R_{\mu_1\ldots\mu_4} = {1\over R_{AdS}^2} \bigl( g_{\mu_1\mu_3}g_{\mu_2\mu_4} - 
g_{\mu_1\mu_4}g_{\mu_2\mu_3}\bigr) \ .  
\fqr 
Within the {\it strong} form of the correspondence the equivalence holds 
at all values of the couplings in \rf{couplings}\footnote{The integral functions 
upon which a correlation function, not restricted by conformal invariance in the 
four-point, may by expanded upon Appell functions from holographic string theory 
\cite{Chalmers:2000vq}; these arise in OPE analysis of conformal field theories 
\cite{Dolan:2000uw} (and references therein)}.  The map between gauge correlations of operators 
${\cal O}_i(x_i)$ on the boundary and holographic string theory in the bulk is 
described in \cite{Witten:1998qj} in which the sources for the composite operators are the 
boundary data of higher-dimensional on-shell string fields $\phi_{o,i}$ via $S_{\rm int} 
= \int d^d{\vec x}~ \phi_{o,i} {\cal O}_i(x_i)$.  An explicit coupling description of 
the genus expansion of the gauge theory through covariantized holographic string scattering 
has been explored in \cite{Chalmers:2000vq}.  

There are three regions in the parameter space of couplings that we explore in the 
gauge theory and IIB string theory: 
\bqr 
& &  
I. ~~~~ \lambda = g_{\rm YM}^2 N_c \; \;\; \; \; {\rm finite} 
 \, \;\;\;\; \;\;\;\;  N_c ~~ {\rm large} 
\cr && 
II. ~~~  \lambda = g_{\rm YM}^2 N_c \;\;\;\; {\rm large} \, \;\;\;\; 
\;\;\;\;g_{\rm YM}^2 = \lambda/N_c ~~ {\rm small} 
\cr && 
III. ~~ \lambda_D = \frac{N^2_c}{\lambda} \,\;\;\;\;\; {\rm finite} 
\, \;\;\;\;\;\;\;\;N_c ~~ {\rm large}  
\label{regimes}
\fqr 
The first limit is the expansion in the planar limit of the gauge theory, according 
to which the graphical expansion has the form $N^{2(1-{\tilde g})}_c F(\lambda)$ at 
string genus ${\tilde g}$; infinite $N_c$ is described by tree-level string scattering in the 
AdS/CFT correspondence.  The instanton corrections have fractional dependence on $N_c$ 
and do not contribute as $N_c\rightarrow\infty$.  The second limit describes strong 
coupling at large $N_c$ described in the gauged supergravity approximation of the IIB 
superstring on $AdS_5\times S^5$ at small curvature according to the AdS/CFT duality.  
The third limit in \rf{regimes} is the expansion in the S-dual variables of the gauge 
theory (and string theory); this expansion in the gauge theory is formulated in terms 
of the S-dual degrees of freedom, i.e. the monopoles and dyons in $\cN=4$ super 
Yang-Mills theory.  The coupling structure has been explored for these limits in 
\cite{Chalmers:2000vq}.  We further describe this latter limit in the following.

S-duality, in either $\cN=4$ super Yang-Mills theory or IIB superstring theory, 
exchanges the couplings through the S-generator as 
\bqr  
\lambda\rightarrow \lambda_D={N^2\over \lambda} \ ,\qquad N_c \rightarrow N_c 
\fqr 
and does not commute with the large $N_c$ planar expansion.  S-duality exchanges 
fundamental fields within $\cN=4$ super Yang-Mills theory to those of the monopole 
and dyon degrees of freedom \cite{Montonen:1977sn} (in the spontaneously broken 
case the theory exhibits the BPS mass formula $m^2=2\vert {\vec n}_e \cdot 
{\vec a} + {\vec n}_m \cdot {\vec a}_d\vert^2$ with $a_d{}^i=\tau^i{}_j a^j$; 
the unbroken limit of the theory contains an infinite number of massless monopoles 
and dyons despite the presence of a local microscopic Lagrangian)\footnote{The dual 
description may be obtained by inverting the spontaneously broken theory in the coupling 
and then taking the vacuum values of the scalar fields to zero.  In this manner the 
spectrum of massive states is well-defined in the local sense.}.  The strongly coupled 
limit of the monopole and dyon formulation in the large $N_c$ limit is the expansion of 
\bqr  
\lambda/N_c^2 \sim ~{\rm small} \qquad N_c \sim ~{\rm large} \ .  
\label{scoupling}
\fqr 
We remind the reader that strongly coupled monopole and dyon formulation 
of $N=4$ super Yang-Mills theory involve the limit in \rf{scoupling}, i.e. 
$g^2\rightarrow 0$, and this expansion describes the planar limit of the gauge 
theory in terms of these soliton degrees of freedom, leading to a different 
expansion in the coupling constants, i.e. $\lambda/N^2_c$ and $N_c$. 

This dual expansion of the gauge theory is examined within the holographic 
anti-de Sitter correspondence in \cite{Chalmers:2000vq}.  The genus truncation 
is related to the existence of this limit of the gauge theory, according to 
which the limit exists, and given the S-duality of the gauge theory implies that 
the maximum number of genus corrections to the order $\alpha'^k$ in the derivative 
expansion is $g_{\rm max}= {1\over 2} (k+2)$ for $k$ even and $g_{\rm min} = 
{1\over 2}(k+1)$; example listings are in Table 1.  The genus truncation directly 
prohibits terms in the scattering that diverge as $N_c$ becomes large at finite dual 
coupling \cite{Chalmers:2000vq}.  This form also agrees with the expansion of modular 
invariant functions in accord with IIB superstring theory with the inclusion of the 
multiplicative product of zero-weight functions $E^{(q,-q)}_s(\tau, \bar\tau)$ for 
$s\geq 3/2$ and real, described in \cite{Chalmers:2000zg}.  
  
\begin{table}  
\caption{
Example contributions to $\Box^k R^4$ arising in string perturbation theory at 
genus $g$.  The asterisk denotes the top $g_{\rm max}$ genus contributions.} 
\begin{center}
\begin{tabular}{|l|l|l|l|l|l|r|} \hline 
{\em } &  $g=0$ & $g=1$ & $g=2$ & $g=3$ & $g=4$ & $\ldots$ \\ \hline 
$R^4$      & $\surd$ & $\surd^\star$ & & & & \\ \hline  
$\Box R^4$ &         &         & & & & \\ \hline  
$\Box^2 R^4$ & $\surd$ &   & $\surd^\star$ & & &  \\ \hline 
$\Box^3 R^4$ & $\surd$ & $\surd$ & $\surd$ & & & \\ \hline 
$\Box^4 R^4$ & $\surd$ & $\surd$ & $\surd$ & $\surd^\star$ & & \\ \hline 
$\Box^5 R^4$ & $\surd$ & $\surd$ & $\surd$ & $\surd$ & & \\ \hline 
$\Box^6 R^4$ & $\surd$ & $\surd$ & $\surd$ & $\surd$ & $\surd^\star$ & \\ \hline
$\cdots$ & & & & & & \\ \hline     
\end{tabular} 
\end{center} 
\end{table} 

Apart from the two dualities, a proof of this property of the expansion of the 
S-matrix might be useful for providing information in the opposite direction, that 
is, to further a proof of S-duality.  At the genus two expansion of the four-point 
graviton scattering amplitude \cite{Iengo:1999qh}, the integrand appears to have an  
$R^4$ eight-derivative term that is in contradiction to explicit two-loop 
supergravity calculations.  This term in string theory must integrate to zero and 
is entirely proportional to the extra superconformal ghost degrees of freedom 
\cite{Friedan:1986ge} (i.e. 
picture structure) that must be inserted in the amplitude calculation at higher 
genera.  This suggests, at $g=2$ and similar analysis at $g>2$ (one limit via factorizing 
multi-genus on products of $g=2$), that a justification of the duality structure 
of the scattering might be found in an appropriate interpretation in the target 
space-time of the world-sheet BRST invariance.  These zero-momentum operators, in 
the field theory limit, could be associated with generating cancellations in the 
target space-time graviton scattering leading to the genus truncation in string 
perturbation theory.  

The genus truncation implied by the two independent dualities of string theory 
in the low-energy scattering implies cancellations of the graviton scattering 
in the maximally extended setting.  No divergence has ever been produced in the 
scattering of maximal supergravity and the four-point graviton function at two-loops 
has been determined \cite{Bern:1998ug}.  This theory may be formulated as the toroidal 
compactification and reduction to massless modes of $\cN=1$ $d=11$ supergravity 
or $\cN=2$ $d=10$ IIB supergravity (the complete reduction of which can explicitly 
be shown to agree with perturbative string theory amplitude calculations up to genus 
two).   

The U-duality group $E_{11-d(11-d)}$, which conjecturally connects the five fundamental 
string theories \cite{Hull:1995ys}, contains an $S$-duality subgroup for 
values of $d$ which inverts fractionally the dilaton coupling 
constant.  We consider the special point of the moduli of the toroidally compactified 
theory when all moduli except for the ten-dimensional dilaton have vanishing 
values (this does not lead to enhanced gauge symmetry as the multiplet containing 
the graviton also contains the vector particles).  In the toroidal limit the 
large $\tau_2$ behavior of the Eisenstein series may be constructed.  The functional 
form of the small moduli contribution is invariant under $SL(2,Z)$ 
transformations and is for large values of $\tau_2$,   
\bqr  
E_s^{E_{11-d(11-d)}} (\tau,\bar\tau) = \Bigl[ \tau_2 V_{10-d}^{4\over 10-d} 
\Bigr]^{ ({s (10-d)\over (d-2)}) } ~ \sum_{(p,q)\neq (0,0)} 
 {\tau_2^s\over \vert p+q\tau\vert^{2s}} + \ldots \ ,
\label{Easymp}
\fqr 
the form of which truncates in every dimension, in the same manner as it does 
in $d=10$ (the volume factor is inert under duality transformations and represents 
the dimensionality of the d-dimensional gravitational coupling constant).  
The form in \rf{Easymp} is due to the $SL(2,Z)$ subgroup of the larger 
U-duality group.  Constructions based upon the generalized Eisenstein series 
follows similarly in the toroidal compactifications (and in fractional dimensions $d$), 
with the difference involving the volume factor of the $10-d$ dimensional tori.  The 
perturbative truncation persists in all of the toroidal compactifications and in every 
dimension.  

\section{String theory to gravity field theory} 
\setcounter{equation}{0}

The agreement of the low-energy field theory modelling of the superstring 
requires an infinite number of cancellations or nullifications at higher 
genus for every $\Box^k$ term in the low-energy expansion.  The genus zero 
and two contributions to the $\Box^2 R^4$ term in the ten-dimensional S-matrix, 
for example, indicates that separately every term  
proportional to $\Box^2$ at $g>2$ has to be zero (every perturbative term at 
different genus differs by powers of $\tau_2^2$).  The same structure occurs 
because of the truncation $g_{\rm max}^k={1\over 2}(k+2)$ and $g_{\rm min}^k= 
{1\over 2} (k+1)$, for even and odd $k$, for the higher derivative terms in 
the expansion of the scattering (and in toroidal compactifications preserving the 
maximal supersymmety).  The two-loop integral form of the four-graviton scattering 
has been computed explicitly within field theory in \cite{Bern:1998ug}, and via 
unitarity constructions to be finite up to five loops in $d=4$ \cite{Bern:1998ug}.  
We describe the continuation of the predictions  of duality at an infinite number of 
loops in this letter, and we compare with the known explicit calculations.  

We analyze in the following the first few terms in the S-matrix expansion 
to outline the cancellations; we discuss $d=10$ with a similar analysis in 
compactified dimensions.  In the IIB supergravity limit of the superstring 
scattering, primitive divergences in $d=10$ of the four-point amplitude at 
one-loop are of the form, with coefficients determined by the domain of 
integration inherited from the string moduli space,  
\bqr 
A_4^{L=1,m=0} \sim \Lambda^2 R^4 + \Box R^4 + \ldots
\fqr 
\bqr 
A_4^{L=1,m\neq 0} \sim  \Box R^4 + \alpha' \Box^2 R^4 + \ldots \ ,
\label{onemassloop}
\fqr 
from an explicit supergravity one-loop calculation or arising from the 
low-energy expansion of the genus one amplitude \cite{Green:1982sw}.  
The $\alpha'$ in \rf{onemassloop} indicates an overall factor of two 
derivatives, a $\Box$ or $s_{ij}$ in momentum space, in comparison to 
the massless modes after they are normalized correctly at the given order.

At two-loops in the supergravity theory, an explicit twelve derivatives may be 
extracted from the loop integration \cite{Bern:1998ug} and the amplitude has the 
generic tensor structure, 
\bqr 
A_4^{L=2,m=0} \sim \Box^2 \left(\Lambda^6 + \Lambda^4 \Box +\ldots \right) R^4  \ .
\fqr 
Explicit string theory calculations at genus two may be computed in the $\alpha' 
\rightarrow 0$ limit \cite{Iengo:1999qh}.  The massive modes of the string contribute 
at an order $\alpha'$ higher in the low-energy limit (as at order $g=0$ and $g=1$), 
and an additional pair of derivatives must also be extracted, 
\bqr 
A_4^{L=2,m\neq 0} \sim \Box^3 R^4 + \ldots  \ . 
\fqr 
This is in agreement with the conjecture for the $\Box^2 R^4$ term in eleven 
dimensions \cite{Green:2000pu} upon dimensional reduction and T-dualized to IIB 
superstring theory in ten dimensions.  

At tree and one-loop level the massive modes explicitly contribute an order higher 
in $\alpha'$, and such a property persists at higher order in the string-inspired 
regulator.  This may also be analyzed via decoupling and unitarity considerations 
(superconformal ghost insertions complicate an analysis directly in the string 
scattering).  Via decoupling, the domain of integration in the generalized Schwinger 
proper-time regulator captures both the UV behavior of both the massless and massive modes.  
For example, at genus one the integration is over the fundamental domain in \rf{modregion}.  
The string theory is not directly an infinite summation over the massive fields, but upon 
comparing the integration of tower of string states, the field theory in the string-inspired 
regulator models the UV behavior.  The string-inspired regulator applied to the massless 
modes reflects the massive modes through the definition associated with the 
integration over the moduli of the world-sheet (the integration region associated 
with integrating over the world-line of a propagating field).   

Multi-loop graviton scattering generically produces two-derivatives 
at every vertex that must be integrated in individual diagrams.  
The perturbative counting arising from the Eisenstein series construction, i.e. 
$g_{\rm min}=0$ to $g_{\rm max}^k$, has the same behavior in the 
field theory by an extraction of an additional four derivatives 
at every successive loop order (as found for example at loop order one 
and two).   We denote the number of cancelled components of the internal 
tensor associated with derivative vertices by $N_L = 4(1+L)$; such a structure 
in Feynman diagrams also arises in $\phi^3$ theory.  In the limit in 
which the ladder diagrams are isolated, and in the conjectured 
construction of \cite{Bern:1998ug}, this property is seen to infinite 
loop order (before integration) for the massless modes, and pure $\cN=8$ 
supergravity at vanishing moduli is perturbatively finite in this regime to all 
orders in four through six dimensions.  The dualities indicate the same 
structure after summing over {\it all} perturbative diagrams at every loop 
order; complete three-loop expressions based on quantizing the $N=8$ 
theory for graviton scattering have not been obtained yet although many 
contributions have been isolated through cut-constructibility in 
\cite{Bern:1998ug}.  Previous work based on the hypothetical existence of an 
unconstrained off-shell superspace indicates cancellations up to seven loops 
\cite{Grisaru:1982zh} (although such a formalism containing only a finite number 
of fields has not been constructed).  These cancellations via the Eisenstein 
formulae occur in every integral dimension; however, the domain of integration 
over the supergravity modes in first quantized form over the world-line parameters 
is unchanged dimensionally. 

The tensor property, i.e. an extraction or nullification of internal loop 
momenta associated with the gravitational couplings in the pattern $4(1+L)$, 
implies that $\cN=8$ supergravity with all moduli tuned to zero except for the 
dilaton coupling is finite in four through six dimensions.  The extension of 
S-duality to general values of dimension generates regulator independence in 
the predictions.  A calculation of the primitive divergence in maximal supergravity 
at three-loops in ten dimensions is useful to further establish the modular 
invariance predictions of S-duality in the superstring truncated to the massless 
modes.  The graviton scattering indicates that this primitive divergence is 
suppressed by four powers and is of order $\Lambda^{10}$, the first cancellation 
in a series related to to the counting of reduced tensor structure $4(1+L)$ at 
$L$ loops.  

Given the two dualities, both S-duality (and U-duality) and the AdS/CFT duality, 
at finite values of couplings the supersymmetric perturbative quantum theory of 
gravity in the maximally supersymmetric case appears finite.  These dualities 
have not been proved;  however, there is evidence for both of these dualities in 
the extension to arbitrary energies and couplings.   

\vskip .3in 
\noindent Acknowledgements 
\vskip .2in  

The work of G.C. is supported in part by the U.S. Department of 
Energy, Division of High Energy Physics, Contract W-31-109-ENG-38. G.C. 
thanks Carlos Wagner for encouragement in this work.

\end{document}